\documentclass[numreferences]{article}
\usepackage{latexsym,amsmath,amsthm,amssymb}

\usepackage{graphicx,epsfig}

\begin{document}

\title{Cosmological Principle and Honeycombs}

\author{C. Criado \thanks{Partially
supported by the Spanish Research Grant FQM-192}\\ Departamento de
Fisica Aplicada I,\\ Universidad de Malaga, 29071 Malaga, Spain
\\ (c\_criado@uma.es) \and N. Alamo
\thanks{Partially supported by the Spanish Research Grant
BFM2001-1825}\\Departamento de Algebra, Geometria y Topologia,\\
Universidad de Malaga, 29071 Malaga, Spain \\
(nieves@agt.cie.uma.es)}

\date{}
\maketitle

\begin{abstract}
We present the possibility that the gravitational growth of
primordial density fluctuations leads to what can be considered a
week version of the cosmological principle. The large scale mass
distribution associated with this principle must have the
geometrical structures known as a regular honeycombs. We give the
most important parameters that characterize the honeycombs
associated with the closed, open, and flat
Friedmann-Lema\^{i}tre-Robertson-Walker models. These parameters
can be used to determine by means of observations which is the
appropriate honeycomb. For each of these honeycombs, and for a
nearly flat universe, we have calculated the probability that a
randomly placed observer could detect the honeycomb as a function
of the density parameters $\Omega_{0}$ and
$\Omega_{\Lambda0}$.\vskip.5cm
 \noindent {\it PACS:\/ }
98.65.Dx, 98.80.-k, 04.20.-q.

\end{abstract}

\section{Introduction}
In recent works (see
Refs.~\cite{aeinasto,beinasto,broadhurst,kirshner,bbattaner}), it
has been speculated that the cosmological large scale matter
distribution may form repetitive structures analogous to the
crystalline ones formed with polyhedra. The geometrical structures
of largest symmetry into which a homogeneous space can be
decomposed are known as regular honeycombs. A regular honeycomb is
a decomposition of the space into congruent regular polyhedra (see
\cite{coxeter,bvinberg}). In Sect.~2 we give a new version of the
cosmological principle that we call the weak cosmological
principle. This version is the most natural way to extend the
cosmological principle to a universe with inhomogeneities. The
geometrical structures that fit it are precisely the regular
honeycombs. Because closed, open, and flat
Friedmann-Lema\^{i}tre-Robertson-Walker (FLRW) models correspond
to the three possible homogeneous spaces: elliptic, hyperbolic,
and Euclidean respectively, we have given the distances, angles
and others characteristic parameters for the regular honeycombs of
these spaces (Sects.~3 and~4). The calculation of the distances
and angles appears in the appendix. Observe that among these three
spaces only the Euclidean one does not have a proper scale. The
other two have a characteristic scale, namely, the curvature
radius (see \cite{thurston}). Thus, in these spaces there are no
homothetic polyhedra and the honeycombs are rigid in the sense
that the size of the basic cell cannot be arbitrary as is the case
for the Euclidean honeycombs. In particular, in these spaces there
cannot be fractal structures because fractals do not have any
characteristic scale. In Sect.~5 we present a possible
interpretation of this geometric scenario in terms of the
cosmological dynamic, by interpreting the spherical and hyperbolic
honeycombs as the pattern of large scale mass distribution. Such
crystalline structures would be the result of the evolution of a
homogeneous one. If the basic cell of the honeycomb lies in the
interior of the particle horizon then the appropriate honeycomb
could be determined by observations, providing a check on the
parameters values that we have calculate in sections~3 and ~4. In
section~6 we have detailed some observational prospect for
detection of the honeycomb structure. The case in which the size
of the basic cell is much smaller than the particle horizon is
commented in subsection~6.1. In subsection~6.2 we have studied the
case that the particle horizon is of the order of the basic cell's
size, and we have calculated, for each regular honeycomb, the
probability that a randomly placed observer could detect the
honeycomb as a function of the density parameters $\Omega_{0}$ and
$\Omega_{\Lambda0}$.

\section{A weak version of the cosmological principle}
The assumption of large-scale homogeneity together with that
 of large-scale isotropy, is called the cosmological
principle (CP) (see, for example, \cite{rindler}). This principle
applies for continuous mass distribution. For the discrete case,
the symmetry of the regular honeycombs is the most natural
definition of discrete homogeneity and isotropy. A regular (or
homogeneous) honeycomb is a decomposition of the space into
congruent regular polyhedra, which are called the cells of the
honeycomb. Any motion that takes a cell into another, takes the
whole honeycomb into itself, i.e., belongs to the group of
symmetries of the honeycomb. The homogeneity corresponds to the
decomposition of the space into regular polyhedra (see
\cite{bvinberg}), and the isotropy corresponds to the symmetry of
regular polyhedra. When all the matter is distributed
homogeneously at the vertices of a honeycomb, we say that it obeys
the discrete cosmological principle (DCP). We can consider another
version of the CP, in which the matter is distributed in a
continuous  way, in the pattern of a honeycomb, with a
hierarchical distribution of matter densities, increasing though
the sequence: interior, faces, edges, and vertices of the basic
cells. In general, if a distribution of matter has the symmetry of
a honeycomb, we say that it obeys the weak cosmological principle
(WCP). It includes the CP when the matter is distributed
homogeneously and isotropically in the basic cell, and also
includes the considered above DCP as a limit case. Another weak
version of the cosmological principle has been considered in
\cite{inoue,roukema}. In that version the universe is locally
homogeneous and isotropic but not necessarily globally homogeneous
and isotropic.

In a space of dimension $3$ with constant curvature, the
honeycombs are classified by means of three integer numbers $\{
p,q,r\}$ called the Schl\"afli symbols, which completely
characterize the honeycomb (see \cite{bvinberg}). Specifically,
$\{ p,q\}$ characterizes the polyhedron which is the basic cell of
the honeycomb, $p$ is the number of vertices (or edges) of each
regular polygon that constitute the faces of a cell, $q$ is the
number of faces (or edges) having a common vertex in each cell,
and $r$ is the number of cells having a common edge. Therefore the
dihedral angle of each cell, $\alpha$, equals $ {2\pi}/{r}$. Note
that because the dihedral angle must be a divisor of $ {2\pi}$ not
all the regular polyhedra can be the cells of a regular honeycomb.
It is easy to see that the number of vertices, V, edges, E, and
faces, F, in a polyhedron with Schl\"afli symbols $\{ p,q\}$ can
be given in terms of $p$, $q$ by
\begin{equation}
V = {{2}\over{1 - q({{1}\over{2}} - {{1}\over {p}})}}, \quad E =
{{q V} \over {2}}, \quad {\hbox{and}} \quad F = {{q V} \over {p}}
\label{eq:siete}.
\end{equation}

\begin{table}
\caption{       Three-dimensional spherical
honeycombs.}\label{Tablauno}
 \centerline{\begin{tabular} {ccccccc}
\hline Name   & Schl\"afli   & $N_0$ & $N_1$ & $N_2$ & $N_3$  &
Basic \\
     &symbol  & & & & &cell   \\
\hline
 $5-cell$  & $\{ 3,3,3\}$ & 5 & 10 & 10 & 5 &tetrahedron\\
$8-cell$   & $\{ 4,3,3\}$ & 16 & 32 & 24 & 8 & cube\\ $16-cell$ &
$\{ 3,3,4\}$ & 8 & 24 & 32 & 16   & tetrahedron  \\ $24-cell$ &
$\{ 3,4,3\}$ & 24 & 96 & 96 & 24   & octahedron \\ $120-cell$ &
$\{ 5,3,3\}$ & 600 & 1200 & 720 & 120  & dodecahedron\\ $600-cell$
& $\{ 3,3,5\}$ & 120 & 720 & 1200 & 600   & tetrahedron\\
 \hline
\end{tabular}}
\end{table}

\section{Honeycombs in the closed FLRW models}

In this case the space-like sections of the universe are three
dimensional spheres, $S^3$. Honeycombs in $S^3$ are in one-to-one
correspondence with regular polyhedra in $R^4$. The correspondence
can be described as follows. The convex hull of a set $M$ is the
minimal convex set containing this set; it is the intersection of
all convex sets containing $M$. Then the convex hull in $R^4$ of
the set of vertices of a honeycomb in $S^3$ is a regular
polyhedron inscribed in $S^3$, and conversely if $P$ is a regular
polyhedron inscribed in $S^3$, then the central projection of its
faces onto $S^3$ forms a honeycomb in $S^3$. From the six regular
polyhedra of $R^4$ (see \cite{coxeter,bvinberg}) we get the
following six regular honeycombs in  $S^3$:

The regular simplex of  $R^4$ with Schl\"afli symbols $\{ 3,3,3\}$
gives the honeycomb 5-cell of $S^3$, which is composed of 5
spherical tetrahedra.

The regular cube of $R^4$ ($\{ 4,3,3\}$) gives the 8-cell of
$S^3$, which is composed of 8 spherical cubes.

 The regular cocube of $R^4$ ($\{ 3,3,4\}$) gives the 16-cell of $S^3$,
which is composed of 8 spherical tetrahedra.

 The regular 24-hedron ($\{ 3,4,3\}$) gives the 24-cell of $S^3$,
which is composed of 24 spherical octahedra.

 The 120-cell honeycomb ($\{ 5,3,3\}$) is composed of 120 spherical
dodecahedra.

 Finally the 600-cell honeycomb ($\{ 3,3,5\}$) is composed of 600 spherical
tetrahedra.

Reversing the order of the Schl\"afli symbols yields the so called
dual honeycombs. The vertices of the honeycomb ${\cal P^*}$, dual
to the honeycomb ${\cal P}$, should be taken as the centres of the
cells of ${\cal P}$. The symmetry groups of ${\cal P}$ and ${\cal
P^*}$ coincide. The honeycombs 8-cell and 16-cell are dual to one
another, and the same holds for the 120-cell and 600-cell. For
symmetric Schl\"afli symbols, dual honeycombs are congruent. This
is the case of the 5-cell and the 24-cell. In our interpretation
of the honeycombs as the patterns of the large scale matter
distribution, if the higher density is at the vertices of a
honeycomb then the lower density is at the vertices of the
corresponding dual honeycomb.

Table~\ref{Tablauno} gives the following characteristics of these
honeycombs: the Schl\"afli symbols  $\{ p,q,r\}$; the number of
vertices,  $ N_0$,  edges, $ N_1$,  faces, $ N_2$, and polyhedra,
$ N_3$. Note that $N_0-N_1+N_2-N_3$ is the Euler characteristic of
$S^3$, so it is zero. In Table~\ref{Tablados} we also give the
distance between adjacent vertices (or edge-length), $d$; the
distance from the centre $C$ of a cell to a vertex $V$ (or
circum-radius), $r_c$; to an edge $E$, $d_E$; to a face $F$ (or
in-radius), $r_i$; and the distance from the centre of a face to a
vertex of that face, $d_F$. The calculation of these parameters is
shown in the appendix.
\begin{table}
\caption{Characteristic parameters associated with the
three-dimensional spherical honeycombs
($R(t)=1$).}\label{Tablados}
\centerline{\begin{tabular}{cccccccccc} \hline Name &$d$& $r_c$
&$d_E$ &$r_i$ &$d_F$ &${\hbox{Vol}}$&
$\rho$\\
 \hline
  $5-cell$&1.8235& 1.3181 &1.1503 & 0.9117 &1.1503  & 3.9478&  0.2533\\
 $8-cell$&1.0472& 1.0472 &0.9553 &  0.7854 &0.7854  &2.4674 &  0.8105\\
 $16-cell$&1.5708& 1.0472  &0.7854&  0.5236 &0.9553  &1.2337  &  0.4023\\
 $24-cell$&1.0472& 0.7854  &0.6155&  0.5236 &0.6155  &0.8224&  1.2158\\
 $120-cell$&0.2709& 0.3881  &0.3648&  0.3141 &0.2318  &0.1644  &  30.3964\\
 $600-cell$&0.6283& 0.3881  &0.2318&  0.1354 &0.3649 &0.0329 &  6.0793\\
\hline
\end{tabular}}
\end{table}
{\setlength\arraycolsep{4pt}

The distances in the R-W spherical space for any cosmic time $t$
are the above multiplied by the expansion function of the
universe, $R(t)$. To obtain the corresponding recessional
velocities we have to multiply the above distances by the Hubble
parameter, $H(t)$.

\begin{table}
\caption{ Other characteristic parameters associated with the
three-dimensional spherical honeycombs.}\label{Tablatres}

\centerline{\begin{tabular} {ccccc} \hline Name    & Dihedral  &
Interior&$C_V$&$E_V$   \\
       & angle & angle &&  \\
\hline $5-cell$      &  $120^o$ & $109.47^o$ & 4 &4  \\
$8-cell$       & $120^o$  & $109.47^o$ & 4 &4 \\
$16-cell$ &  $90^o$  & $90^o$ & 8 & 6 \\
$24-cell$ & $120^o$  & $70.53^o$  & 6 & 8 \\
$120-cell$       & $120^o$ &$109.47^o$& 4 & 4\\
$600-cell$ & $72^o$   &$63.43^o$& 20 & 12\\
\hline
\end{tabular}}
\end{table}

We can also obtain the volume of a cell, ${\hbox{Vol}}$, as the
quotient of the volume of $S^3$, $2\pi^2R(t)^3$, and the number
$N_3$ of cells of the honeycomb. The density of vertices $\rho$ is
the quotient of $N_0$ and the volume of $S^3$. In
Table~\ref{Tablados} we list the values of the volume
${\hbox{Vol}}$ and $\rho$ for the six honeycombs considered. Other
interesting parameters of the honeycombs are: the number of edges
that share a vertex, $E_V$, which is given by $E_V = 2N_1/N_0$ and
the number, $C_V$,  of cells that share a vertex, which is given
by $C_V = rE_V/q$. This number corresponds also to the number of
vertices of basic cell of the dual honeycomb, and thus is given by
$C_V=1/{(1-q(\frac{1}{2}-\frac{1}{r}))}$. In table~\ref{Tablatres}
we list the values of these parameters, as well as the dihedral
angle, and the interior angle $\phi$ of the polygons constituting
the faces of each cell.

\section{Honeycombs in the open and flat FLRW models}
The space-like sections of the universe, in the open FLRW model,
are 3-dimensional spaces of constant negative curvature, and these
spaces are isomorphic to ${H^3}$, the hyperbolic space (or
Lobachevskij space) of dimension 3. If we restrict ourselves to
honeycombs with bounded cells it follows that there are only four
regular honeycombs in $H^3$ \cite{bvinberg}( see the appendix).
Their Schl\"afli symbols and dihedral and interior angles are
listed in table~\ref{Tablacuatro}. We have also
 calculated the characteristic distances and angles of the basic cell of
these honeycombs, as well as the volume of the basic cell, the
number, $E_V$, of edges that share a vertex, the number, $C_V$, of
cells that share a vertex, the volume of the basic cell, Vol, and
the density, $\rho$. Table~\ref{Tablacuatro} and~\ref{Tablacinco}
give the values of all these parameters. Note that the honeycombs
$d_{90}$ and $c_{72}$ are dual to one another, and that $d_{120}$
and $d_{72}$ are self-dual.

\begin{table}
\caption{Characteristic parameters associated with the
three-dimensional bounded honeycombs of the open R-W space
}\label{Tablacuatro} \centerline{\begin{tabular} {ccccccc} \hline
Name   & Schl\"afli & Basic   & Dihedral  & Interior &$C_V$&$E_V$
\\
     &symbol   &cell   & angle & angle&&   \\
\hline $i_{120}$ & $\{ 3,5,3\}$& icosahedron& $120^o$ & $41.81^o$ & 12 &20\\
 $d_{90}$ & $\{ 5,3,4\}$& dodecahedron & $90^o$& $90^o$ & 8 &6 \\
 $c_{72}$& $\{4,3,5\}$   & cube & $72^o$  & $63.43^o$ & 20 &12 \\
 $d_{72}$ & $\{5,3,5\}$& dodecahedron & $72^o$  & $63.43^o$ & 20 &12 \\
\hline
\end{tabular}}
\end{table}

\begin{table}
\caption{Other characteristic parameters associated with the
three-dimensional bounded honeycombs of the open R-W space
($R(t)=1$).}\label{Tablacinco}
\centerline{\begin{tabular}{cccccccccc} \hline Name &$d$& $r_c$
&$d_E$ &$r_i$ &$d_F$ &${\hbox{Vol}}$& $\rho$\\
 \hline
 $i_{120}$&1.7366 &1.3826 & 0.9726& 0.8683& 0.9727& 4.6860  & 0.2134\\
 $d_{90}$&1.0613 &1.2265 &1.0613 &  0.8085& 0.8425 & 4.3062    & 0.5806\\
 $c_{72}$&1.6169 &1.2265 &0.8425 &  0.5306 & 1.0613& 1.7225  & 0.2322\\
 $d_{72}$&1.9927 &1.9028 &1.4391 &  0.9964 &1.4321 & 11.1991   & 0.0893\\
\hline
\end{tabular}}
\end{table}

The flat universe corresponds to the Euclidean tridimensional
space. The regular polyhedra of this space are the five platonic
polyhedra. Among these polyhedra only the cube has the dihedral
angle divisor of $2\pi$. Thus the only possible regular honeycomb
is formed by cubes. Its Schl\"afli symbols are $\{4,3,4\}$, the
dihedral and the interior angles are both of $90^o$, $C_V=8$, and
$E_V=6$. But because in a flat space there are not a proper
length, the basic cube can be of any size and there is no
characteristic distance.

We can also consider whether there are honeycombs structures for
models of the universe with local constant curvature but with
topology different to the usual one. Models of universes of this
type have been considered; see
\cite{cornish-spergel-starkman,weeks,cornish-weeks}. See also
\cite{cornish-spergel-starkman-komatsu} for a recent result that
constrain the possible topology of these spaces. Special attention
has been paid to the locally flat and the locally hyperbolic ones.
The reason is that one can then have universes that are compact
and flat, and universes that are compact and have negative
constant curvature respectively. One which is very popular is
known as the Seifert-Weber dodecahedral space (see
\cite{thurston,thurston-weeks}). This space is obtained from the
above $d_{72}$ hyperbolic honeycomb. To construct this space we
have to glue together the opposite faces of the basic dodecahedron
using a clockwise twist of $3/10$ of a revolution. Another
example, this one with positive constant curvature, is the
Poincar\'e dodecahedral space. This space is associated with the
120-cell spherical honeycomb. To obtain it, opposite faces of the
basic honeycomb are glued together using this time a twist of
$1/10$ of a revolution. For the flat space, identification of the
opposite faces of a cube gives the 3-torus, which is a compact
flat model of the universe. Only this last space, among all spaces
with non trivial topology, admits a regular honeycomb structure.
This is because for constant non zero curvature there are not two
basic polyhedra whose distances between vertices, are such that
one is a divisor of the other.

\textit{Comment.} The relation of the above regular honeycombs and
the multiply connected spherical orientable spaces (see
\cite{lachieze-rey-luminet,inoue1,levin}) is as follows: to obtain
a spherical orientable 3-manifold by identifying the faces of a
platonic polyhedron $\Sigma$, the polyhedron must obey two
conditions (see \cite{everitt}): (1) the dihedral angle must be a
submultiple of $2\pi$, say $2\pi/r$, and (2) the number of edges
of $\Sigma$ must be divisible by r. By definition, the basic
polyhedron cell of any honeycomb obeys (1), but there are two
spherical honeycomb, the 16-cell and the 600-cell, that do not
satisfy (2). With the remaining four spherical honeycombs we can
associate globally homogeneous spherical 3-manifolds. These
manifolds are single action spherical manifolds. The single action
spherical manifolds are those for which the members of a subgroup
$R$ of $S^3$ act as pure right-handed Clifford translations (see
\cite{lehoucq}). With the honeycomb 5-cell, whose basic cell is a
tetrahedron, we can associate the lens space $L(5,1)$, which is
the single action manifold associate to the cyclic group $Z_{5}$.
With the honeycomb 8-cell, whose basic cell is a cube, we can
associate the
 Montesinos's quaternionic  space, which is a prism manifold
 associate to the binary dihedral group $D^{*}_{2}$. With the
 honeycomb 24-cell, whose basic cell is an octahedron, we can
 associate the Montesinos's octahedral space, which is the single action
 manifold associate to the binary tetrahedral group $T^{*}$.
 Finally, with the honeycomb 120-cell, whose basic cell is a
 dodecahedron, we can associate the above-mentioned Poincar\'e
 dodecahedral space, which is the single action manifold associate
 to the binary icosahedral group $I^{*}$. These associations provide a
 way to show that $S^{3}$ is a covering space of the above
 manifolds.

 Among the four hyperbolic honeycombs only the $i_{120}$ and the
 $d_{72}$ obey the above condition (2). By identification of
 opposites faces of the basic icosahedron cell of $i_{120}$ we get
a hyperbolic compact manifold, the 3-torus $T^{3}$. The
  basic dodecahedron cell of the honeycomb $d_{72}$,
   gives rise by identification of opposites faces,
 to the above-mentioned Seifert-Weber dodecahedral space. Finally,
 \cite{everitt} gives other possible manifolds associated
 with these honeycombs.

\section{Interpretation in terms of standard FLRW cosmology}

Analysis of the power spectrum of density perturbations and the
correlation function have shown that galaxies appear to be
gathered into immense sheets and filaments surrounding very large
voids (see Refs.~\cite{aeinasto,beinasto,broadhurst}). The most
symmetric distribution of matter, after the homogeneous and
isotropic one, are those associated with the honeycomb structures.
These structures give the most natural generalization of the
cosmological principle (CP). We have named this generalization
weak cosmological principle (WCP). Then we propose that the large
scale structure of the universe could have the structure of a
honeycomb.

We have seen that there are eleven suitable honeycombs, six
corresponding to a closed universe, four to an open one, and one
to a flat one. We have calculated the different parameters that
characterize these honeycombs.

The model that we propose is very speculative, but we think that
it could be useful in looking for new ways to interpret the
inhomogeneities that has been discovered on large cosmological
scales. To make this scenario feasible we have to assume that,
initially, there was a homogeneous and isotropic distribution of
dark matter or of some other non observable kind of matter. We
accept also that inhomogeneities with higher energy density than
the mean, formed during the cosmic evolution, are distributed in
the most homogeneous and isotropic manner possible, which we
assume to have the honeycomb structure. From this, we can
speculate with the fact that the visible matter is concentrated in
these inhomogeneities of higher density, with a hierarchical
distribution of densities, increasing through the sequence:
interior, faces, edges, and vertices of the basic cells.

The above symmetric distribution may be considered as the limit
attractor of the less symmetrical present distribution consisting
of a huge net of filaments made up of clusters of galaxies. This
net would evolve seeking the stability associated with the
symmetry of any of the above-described honeycombs. At the present
time we could be just in the phase transition that goes from a
more or less homogeneous distribution to a crystalline one. We do
not know what the precise dynamic governing the above process
might be. Presumably, it would be a very complex one, with the
extragalactic magnetic field as a principal actor. It is possible
that the seeds of these structures were generated in the first
moments after the big bang, perhaps before the inflation due to
the strong magnetic fields generated by the turbulence of the
charged plasma (see \cite{bbattaner,abattaner,florido,cbattaner}).
If this were the case, there would be a suppression of
cosmological density fluctuations on scales beyond the size of the
basic cell, similar to what happens in small universes models (see
\cite{luminet,lehoucq}). Then the honeycomb models could also
explain the existence of a cut-off in the cosmic microwave
background (CMB) angular power spectrum on large angular scales
(see \cite{bennett}).

\section{Observational prospect}

\subsection{ The case in which the size of the basic cell of the honeycomb is much smaller than the
particle horizon}
 If the size of the basic cell of the honeycomb
is much smaller than the particle horizon we could verify the
correctness of the above model. A possible way would be to study
the distribution of high redshifts, z, in any direction. They
should exhibit peaks with periodic separations in $\log (1+z)$.
The period should depend on the periodic structure of the
honeycomb and, therefore, on the observational direction.
Sufficient observations of this kind would enable the
determination of the appropriate honeycomb. Regularities of this
type has been reported by Broadhurst et al (see
Refs.~\cite{broadhurst}). They found that in regions of small area
around the northern and southern galactic polar caps, the high and
low density alternate with a rather constant step of $128 h^{-1}
Mpc$. In other directions the regularity is much less pronounced.

Another possible observational parameter could be the number of
filaments that converge on a supercluster. The open space
honeycombs only admit 6, 12, or 20 filaments; the possibilities
for the closed space are 4, 6, 8 and 12; the flat one only admit 6
filaments (see tables~\ref{Tablatres} and \ref{Tablacuatro}).

Once we know the honeycomb we can use its characteristic distances
 to determine the present curvature of the universe $R_0$.
The value of $R_0$ may then be used to sharpen the value of the
density parameter, $\Omega_{tot}$, (it can be calculated from
$\Omega_{tot}=1-kc^2/(R_0H_0)^2$, $k=1,-1$ for the closed and open
case respectively), as well as other cosmological parameters.

We can look also for observable effects of these structures on the
gravitational waves, analogous to the x-rays diffraction on
crystals. Another possible observational fact is the lens effect
of these periodic structures on electromagnetic waves.

\subsection{The case where the particle
horizon is of the order of the basic cell's size}

If the particle horizon is of the order of the basic cell's size,
we may observe only a part of that basic cell, but the data ratio
between the characteristic distances of the cell as well as the
values of $E_V$ and $C_V$ can be enough to determinate which is
the appropriate honeycomb. To this end, it would be important that
we can observe at least one vertex of the honeycomb, because in
that case we can observe $E_{V}$, $C_{V}$ , and the dihedral and
interior angles, and if these values are the given in the above
tables then we will have evidence that we are in a honeycomb. Also
these values will be enough to determinate the appropriate
honeycomb in all the cases except for the pair of spherical
honeycombs the $5-cell$ and $8-cell$, and the hyperbolic $c_{72}$
and $d_{72}$, for which the values of these four parameters
coincide. To discriminate between these cases, we must use the
observable distances to the vertex, edges, and faces to
reconstruct the basic cell. Now we will calculate for each of the
honeycombs, the probability that a randomly placed observer can
detect a vertex of the honeycomb. This probability, $p_r$, will
depend on the considered horizon radius $r_h$.

In the following we assume that the universe can be described by
the R-W metric, and that the matter is made up from dust of
density $\rho_m$ and a cosmological constant $\Lambda$. The
Friedmann equation is then given by:

\begin{equation}
H^2 ={{{8} \pi G \rho_m}\over {3}}- {{k c^2}\over
{R^2}}+{{\Lambda}\over {3}}
\label{eq:nueve},
\end{equation}
where $H=\dot{R}/R$ is the Hubble parameter, $G$ is the Newton's
constant, and $k=1,0,-1$ for an open, flat, and closed universe
respectively.

Moreover, we have that $\rho_m=(R/R_0)^3 \rho_{m0}$, and the
red-shift $z$, is given by $z=R_0/R-1$, where the subscript $0$
denote evaluation at the present time.

The RW metric gives $dr=(1/R) c dt$ for the photon equation.
Integrating this equation, and taking into account the above
relations, we can find (see \cite{peacock}) that the comoving
distance $r(z)$, run over by a photon as function of the red-shift
$z$, is given by:

\begin{equation}
r(z) =\sqrt{|1-\Omega_{tot}|} \int_{0}^{z}
(\Omega_{\Lambda0}+(1-\Omega_{tot})(x+1)^2+
\Omega_{0}(x+1)^3)^{-{1/2}} dx
 \label{eq:diez},
\end{equation}
where $\Omega_{0}$, $\Omega_{\Lambda0}$, and $\Omega_{tot}$ are
the density parameters given by $\Omega_{0}=\frac{8 \pi G
\rho_{m0}}{3 {H_0}^2}$, $\Omega_{\Lambda0}=\frac{\Lambda c^2}{3
H^2}$, and $\Omega_{tot}=\Omega_{0}+\Omega_{\Lambda 0}$. The above
distance, $r(z)$, is given in units of the curvature radius $R_0$.
The horizon radius, $r_h$, corresponds to $z=\infty$; the last
scattering surface radius, $r_{LSS}$, associated with the cosmic
microwave background (CMB), corresponds to $z\approx 1100$; and
for the quasars  and the clusters of galaxies, we could take
red-shift cut-offs of $z\approx6$ and $z\approx1$ respectively.
The probability $p_r$ associated with any of these radius $r(z)$,
is given by the fraction of the basic cell volume in which the
distance to a vertex is smaller than  $r(z)$, that is :

\begin{figure}
\centerline{\includegraphics[width=18pc]{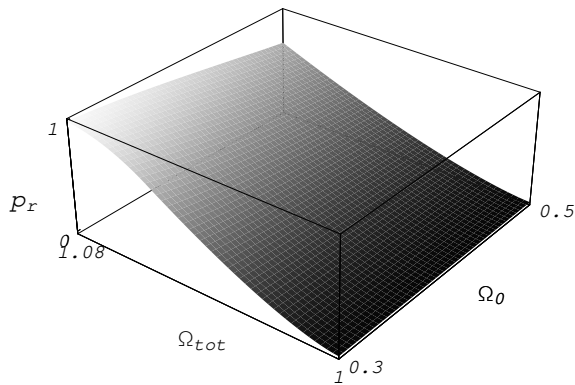}}
\caption{Probability, $p_r$, that a randomly located observer can
detect the spherical honeycomb $16-cell$, for
$0.3<\Omega_{0}<0.5$, $1<\Omega_{tot}<1.08$ and $z=\infty$.}
\label{Figtwo}
\end{figure}

\begin{equation}
p_r=\frac{V_r}{V}
 \label{eq:once},
\end{equation}
where $V_r$ is the volume of the region of the basic cell such
that the distances from its points to a vertex are less than
$r(z)$, and $V$ is the volume of the basic cell of the honeycomb.
If $r(z)>r_c$ then $V_r=V$, and $p_r=1$. If $2r(z)<$ edge-length
$d$, the spheres with centres at the vertices and radius $r(z)$ do
not intersect, and $V_r$ is given by:

\begin{equation}
V_r=V\frac{V_s(r(z))}{C_V}
 \label{eq:doce},
\end{equation}
where ${V_s}(r(z))$ is the volume of the sphere of radius $r(z)$,
$C_V=\frac{2}{1-q(\frac{1}{2}-\frac{1}{r})}$ is the number of
cells around a vertex, and
$V=\frac{2}{1-q(\frac{1}{2}-\frac{1}{p})}$ is the number of
vertices of a cell. The volume of a sphere of radius $r$ in $S^3$
and in $\Pi^3$ is given respectively by:

\begin{equation}
V_s(r)=\pi(2 r-\sin 2r), \quad    V_s(r)=\pi(2 r-\sinh 2r)
 \label{eq:trece},
\end{equation}
where we have taken the curvature radius $R=1$.

If $d/2<r(z)<$ distance from the centre of a face to a vertex of
that face $d_F$, there are no common points to more than two
spheres. The volume of the region of the intersection of two
spheres is the double of the volume of the spherical cup, $V_c$
corresponding to the height $r(z)-d/2$, multiplied by the number
of edges of a cell, $E$, and divided by the number of cells with a
common edge, $r$. Thus we have:

\begin{equation}
V_r=\frac{V}{C_V} V_s(r)-\frac{E}{r} 2 V_c(r-d/2)
 \label{eq:catorce}
\end{equation}

The volume $ V_c(r-d/2)$ equals $V_s(r)/2$ minus the volume of the
spherical segment of height $d/2$, $V_{seg}(d/2)$. But
$V_{seg}(h)=\int_{0}^h A(y)dx$, where $A(y)$ is the area of the
circle of radius $y$. Taking into account that  for the spherical
and hyperbolic cases we have respectively: $A(y)=\pi\sin ^2{y}$,
$\cos {r}=\cos{y}\cos{x}$, and  $A(y)=\pi\sinh ^2{y}$,
$\cosh{r}=\cosh{y}\cosh{x}$, we obtain, for the spherical and
hyperbolic segment volume respectively,

\begin{equation}
V_{seg}(h)=(h-\cos^2{r}\tan{h}),\quad
V_{seg}(h)=(h-\cosh^2{r}\tanh{h}).
 \label{eq:quince}
\end{equation}

With the above expressions we can calculate the probability $p_r$
in all the cases except when $d_F<r(z)<r_c$. In this case we can
approximate the value of $p_r$ by interpolation. Using
equation~\ref{eq:diez} we can express $p_r$ as a function of the
density parameters $\Omega_{0}$, and $\Omega_{tot}$. As an
example, we have shown in Fig.~\ref{Figtwo} the probability for a
randomly located observer of detecting the spherical honeycomb
$16-cell$, for $0.3<\Omega_{0}<0.5$, $1<\Omega_{tot}<1.08$ and
$z=\infty$. Note that if $\Omega_{tot}\rightarrow 0$, then
$r(z)\rightarrow 0$, and the probability $p_r$ also goes to $0$.

\begin{table}
\caption{Probability that a randomly located observer detect a
given honeycomb, for red-shift $z=1,6,1100,\infty$. For the
honeycombs of the closed universe we have taken
$(\Omega_{0},\Omega_{tot})=(0.3,1.03)$, and for the open universe
$(\Omega_{0},\Omega_{tot})=(0.3,0.95)$.}\label{Tablaseis}
\centerline{\begin{tabular}{ccccc} \hline Honeycomb &$z=1$& $z=6$
&$z=1100$ &$z=\infty$ \\
 \hline
 \hline
 $r(z)(0.3,1.03)$&0.135& 0.337 &0.558 & 0.576\\
 \hline
 $5-cell$&0.003& 0.040 &0.173 & 0.190 \\
 $8-cell$&0.008& 0.127 &0.546 &  0.590 \\
 $16-cell$&0.004& 0.064  &0.277&  0.304\\
 $24-cell$&0.012& 0.191  &0.809 &0.860 \\
 $120-cell$&0.311& -  &1&  1  \\
 $600-cell$&0.062& 0.913  &1&  1\\
\hline \hline
$r(z)(0.3,0.95)$&0.170& 0.423 &0.706 & 0.729\\
 \hline
 $i_{120}$&0.004 &0.070 & 0.347& 0.386\\
 $d_{90}$&0.012 &0.191&0.740 &  0.779\\
 $c_{72}$&0.005 &0.076 &0.374 &  0.420 \\
 $d_{72}$&0.002 &0.029 &0.145 &  0.161 \\
\hline
\end{tabular}}
\end{table}

Table~\ref{Tablaseis} gives the probability of detecting any of
the regular honeycombs considered above for $z=1,6,1100,\infty$.
For the honeycombs of the closed universe we have taken
$(\Omega_{0},\Omega_{tot})=(0.3,1.03)$, and for the open universe
$(\Omega_{0},\Omega_{tot})=(0.3,0.95)$. These values of the
density parameters are in the range, $0.9<\Omega_{tot}<1.1$, of
the nearly flat universes that have been given by recent
observations \cite{sievers}. The value omitted in the table
corresponds to a value of $r(z)$ such that $d_F<r(z)<r_c$, and, as
it have been pointed out previously, it can not be calculated with
the above procedure. Notice that in the closed universe the higher
probabilities correspond to the $120-cell$ and $600-cell$
honeycombs, and to $d_{90}$ in the open case.

\begin{figure}
\centerline{\includegraphics[width=30pc]{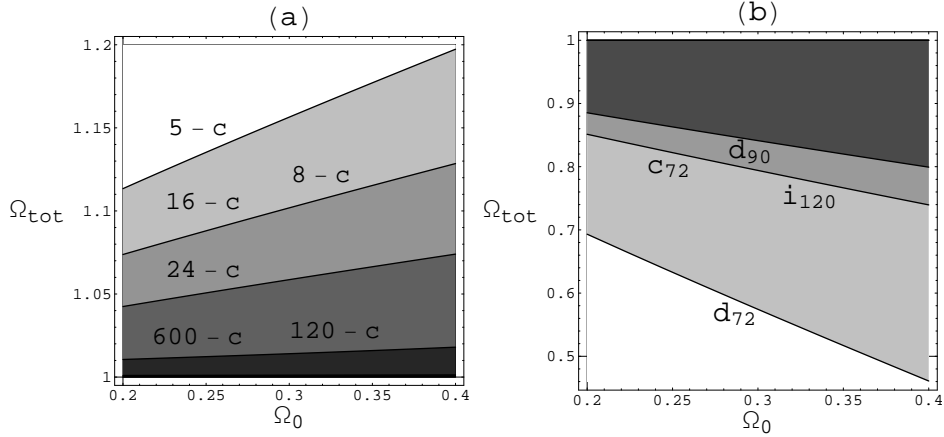}}
\caption{Region of the $(\Omega_{0},\Omega_{tot})$ plane where
$r_{LSS}$ is greater than the circum-radius, $r_c$, that is, the
region on which the probability of detecting the honeycomb is $1$.
Fig.~\ref{Figthree}$(a)$ shows these regions for the spherical
honeycombs. The points in which $r_{LSS} < r_c$ correspond to the
region above the marked line ($r_{LSS} = r_c$). In the same way
Fig.~\ref{Figthree}$(b)$ gives these regions for the hyperbolic
honeycombs. In this case the points with $r_{LSS} < r_c$ are in
the regions below the marked lines.} \label{Figthree}
\end{figure}

We have also calculated for each regular honeycomb the region of
the $(\Omega_{0},\Omega_{tot})$ plane, where $r_{LSS}$ is greater
than the circum-radius, $r_c$, that is, the region on which the
probability of detecting the honeycomb is $1$.
Fig.~\ref{Figthree}$(a)$ shows these regions for the spherical
honeycombs, and Fig.~\ref{Figthree}$(b)$ gives these regions for
the hyperbolic honeycombs. Observe that the honeycombs that are
easier to detect in a nearly flat universe are the spherical
$120-cell$ and $600-cell$.

\section{Summary}

In this article, we have considered the possibility that the
gravitational growth of primordial density fluctuations leads to
what can be considered a week version of the cosmological
principle, for which the large scale matter distribution has the
pattern of a regular honeycomb. In a recently published paper (see
\cite{criado-alamo}) we had studied the honeycombs in the space of
relativistic velocities and in the Milne cosmological model. In
both cases the honeycombs were the hyperbolic ones. In that paper
we advanced some of the ideas of this one.

 There are $6$ regular honeycombs associated with the closed FLRW universe, and $4$
with bounded cells, to the open case. We have calculated the most
important parameters characterizing these honeycombs.

 We have also given some observational prospect for detecting the
 honeycomb. Moreover, we have calculated, for each honeycomb, and for a
nearly flat universe, the probability that a randomly
 placed observer could detect the honeycomb as a function of the
density parameters $\Omega_{0}$ and $\Omega_{\Lambda0}$.

\section{Appendix}

In this appendix we will calculate the characteristic distances
and angles of the regular honeycombs considered in this paper. To
calculate these distances as functions of the Schl\"afli symbols
$\{ p,q,r\}$ of the honeycomb we proceed as follows. First, we
decompose each polyhedron into $F$ identical pyramids with the
apex in the centre of the polyhedron. Each of these pyramids is
then decomposed into $2p$ double-rectangular tetrahedra by
dropping perpendicular lines from the apex onto the faces and onto
the lines bounding the faces. The vertices of this tetrahedron
are: the centre of the cell, $P_3$, the centre of a face, $P_2$,
the centre of an edge, $P_1$, and a vertex of the cell, $P_0$ (see
Fig.~\ref{Figone}). We recall that a tetrahedron $P_0P_1P_2P_3$ is
said to be double-rectangular  if its edge $P_3P_2$ is orthogonal
to the face $P_0P_1P_2$ and its edge $P_1P_0$ is orthogonal to the
face $P_1P_2P_3$. Thus, three out of the six dihedral angles are
right angles. Thus, the double-rectangular tetrahedron is
determined by its dihedral angles $\alpha$, $\beta$, and $\gamma$
corresponding to the edges, $a=P_3P_2$, $b=P_3P_0$, and $c=P_1P_0$
respectively. Then using spherical trigonometry we have (see
\cite{bvinberg})

\begin{figure}
\centerline{\includegraphics[width=12pc]{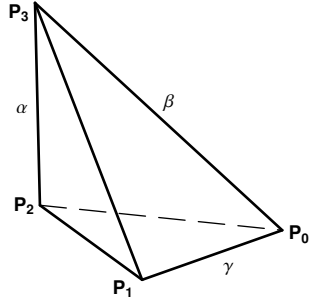}} \caption{One
of the $2pF$ identical double-rectangular tetrahedra which any
regular polyhedron is decomposed into.} \label{Figone}
\end{figure}

\begin{equation}
\tan a \tan \alpha = \tan b \tan ({{\pi}\over {2}} - \beta) = \tan
c \tan \gamma =
 {{\sqrt {\Delta}}\over{\cos \alpha \cos \gamma}}
\label{eq:ocho},
\end{equation}
where $\Delta = \sin^2 \alpha  \sin ^2 \gamma - \cos^2 \beta$.

By definition of $p$, $q$, and $r$ it follows that $\alpha =
{{\pi}/{p}}$,   $\beta = {{\pi}/{q}}$, and $\gamma = {{\pi}/{r}}$.
Therefore, by substituting these values in Eq.~(\ref{eq:ocho}) we
obtain $a$, $b$, and $c$. The above defined characteristic
distances of a honeycomb are then given by: $d = 2c$, $r_c = b$,
$d_E = \arg\sin ({{\sin a}/{\sin \gamma}})$, $r_i = a$, and $d_E =
\arg\sin ({{\sin c}/{\sin \alpha}})$ (see Table~\ref{Tablados}).

Moreover, we can get the interior angle $\phi$ of the polygons
constituting the faces of each cell by solving the hyperbolic
triangle $P_0P_1P_2$. In fact, we have $\sin {{\phi}\over {2}} =
{{\cos \alpha}/ {\cos c}}$. In this way we have obtained the
values of $\phi$ in Table~\ref{Tablatres}.

The space-like sections of the universe, in the open FLRW model,
are 3-dimensional spaces of constant negative curvature, and these
spaces are isomorphic to ${H^3}$. We have followed Vinberg and
Shvartsman \cite{bvinberg} classification of hyperbolic
honeycombs, which does not include as honeycombs those with cells
inscribed in horospheres instead of finite spheres. As in the
spherical case, for a honeycomb with Schl\"afli symbols $\{
p,q,r\}$ the dihedral angle of each cell, $\alpha$, equals $
{2\pi}/{r}$, but in the hyperbolic case $\alpha$ has the
restriction $\alpha_{\mathrm{min}} \leq \alpha <
\alpha_{\mathrm{Euc}}$, where $\alpha_{\mathrm{min}}$ is the
minimal possible dihedral angle in such a regular polyhedron in
the hyperbolic space, and $\alpha_{\mathrm{Euc}}$ is the dihedral
angle of the corresponding polyhedron in the Euclidean space. From
this fact and if we restrict ourselves to honeycombs with bounded
cells it follows that there are only four regular honeycombs in
$H^3$ \cite{bvinberg}. Their Schl\"afli symbols, dihedral, and
interior angles are listed in Table~\ref{Tablacuatro}. In our
paper \cite{criado-alamo}, we
 calculated the characteristic distances and angles of the basic cell of
these honeycombs, as well as the volume of the basic cell, the
number, $E_V$, of edges that share a vertex, the number, $C_V$, of
cells that share a vertex, the volume of the basic cell, Vol, and
the density, $\rho$. Tables~\ref{Tablacuatro} and~\ref{Tablacinco}
give the values of all these parameters.

\end{document}